\documentclass[prd,twocolumn,superscriptaddress,aps]{revtex4-1}

\usepackage{ae}
\usepackage{bm}
\usepackage{xcolor}
\usepackage{amsmath}
\usepackage{amssymb}
\usepackage{amsfonts}
\usepackage{graphicx}
\usepackage{slashed}
\usepackage{wasysym}
\usepackage{units}
\usepackage{ulem}
\usepackage{upgreek}

\newcommand{\Eqref}[1]{Eq.~\eqref{#1}}

\makeatletter
\def\fps@figure{ht}
\def\fps@table{ht}
\makeatother

\allowdisplaybreaks

\begin{document}

\setlength{\unitlength}{1mm}

\title{Exploring new states of matter with a photonic emulator}

\author{Felix Karbstein}\email{felix.karbstein@uni-jena.de}
\affiliation{Helmholtz-Institut Jena, Fr\"obelstieg 3, 07743 Jena, Germany}
\affiliation{GSI Helmholtzzentrum f\"ur Schwerionenforschung, Planckstra\ss e 1, 64291 Darmstadt, Germany}
\affiliation{Theoretisch-Physikalisches Institut, Abbe Center of Photonics, Friedrich-Schiller-Universit\"at Jena, Max-Wien-Platz 1, 07743 Jena, Germany}
\author{Simon St\"utzer}
\affiliation{Institut f\"ur Angewandte Physik, Friedrich-Schiller-Universit\"at Jena, Max-Wien-Platz 1, 07743 Jena, Germany}
\author{Holger Gies}
\affiliation{Helmholtz-Institut Jena, Fr\"obelstieg 3, 07743 Jena, Germany}
\affiliation{Theoretisch-Physikalisches Institut, Abbe Center of Photonics, Friedrich-Schiller-Universit\"at Jena, Max-Wien-Platz 1, 07743 Jena, Germany}
\author{Alexander Szameit}
\affiliation{Institut f\"ur Angewandte Physik, Friedrich-Schiller-Universit\"at Jena, Max-Wien-Platz 1, 07743 Jena, Germany}
\affiliation{Institute for Physics, University of Rostock, Albert-Einstein-Str. 23, 18059 Rostock, Germany}
\affiliation{Department of Life, Light $\&$ Matter, University of Rostock, Albert-Einstein-Str. 25, 18059 Rostock, Germany}

\date{\today}

\begin{abstract}
We implement the equation of motion of the large-$N$ Gross-Neveu model from strong interaction physics in photonic waveguide arrays and study one of its paradigmatic multi-fermion bound state solutions in an optical experiment. The present study constitutes an important first step towards waveguide-based simulations of phenomena relevant for high-energy physics.
\end{abstract}

\maketitle

In recent years, earlier speculations about the existence of new states of fermionic matter in the form of inhomogeneous phases where translational invariance is spontaneously broken have turned into a firm theoretical prediction for certain low-dimensional, exactly solvable field theories such as the Gross-Neveu model \cite{Gross:1974,Thies:2003br,Schnetz:2004}. Similar phenomena have been predicted and extensively studied in a wide variety of research fields, ranging from condensed-matter systems (so-called FFLO phases in superconductors in large magnetic fields) \cite{Fulde:1964zz,Larkin:1964wok}, via ultracold atomic gases, nuclear physics, and the interior of neutron stars, to quark matter at highest densities; see \cite{Casalbuoni:2003wh,Thies:2006ti,Anglani:2013gfu} for reviews. While exact theoretical solutions are now available for 1D models much less is known for the relevant cases of 2D layered structures or in full 3D. This is not merely due to technicalities, but due to the fact that a deeper understanding of the Dirac equation in inhomogeneous phases is conceptually lacking; e.g. higher-dimensional analogues of the 1D Peierls instability are still searched for \cite{Park:2019}. Moreover, to date fermionic matter characterized by a spontaneously broken translational symmetry is extremely difficult to realize in an experiment. However, testing such phenomena experimentally would provide unprecedented insights into the very foundations of theory and trigger conceptually new theoretical approaches for the description of these systems.

Modern optics and photonics is driven by the fact that photons can be coherently controlled in space and time at the highest precision level. This goes hand in hand with recent developments in precision fabrication and design of optical systems. A prominent example is given by waveguide arrays \cite{Longhi:2009} that can be designed as photonic analogues of systems governed by complex wave equations, including even quantum mechanical equations such as the relativistic Dirac equation. Appropriately designed photonic waveguide arrays with alternating refractive indices of adjacent lattice sites have already proved their capability to emulate a variety of relativistic phenomena in a wide range of parameter regimes, including Zitterbewegung \cite{Dreisow:2010}, pair creation \cite{Dreisow:2012}, particles with random mass \cite{Keil:2013}, ultra-strong magnetic fields \cite{Rechtsman:2013} and even tachyons \cite{Song:2020}. Since the Dirac equation governs the dynamics of almost all known matter particles in the universe at the microscopic level, photonics has the potential to explore states of fermionic matter in an unprecedented way.

In this work, we present a photonic emulator for the physics of relativistic fermion systems and apply it to the massless Gross-Neveu model in the large-$N$ limit \cite{Gross:1974}. Remarkably, at low temperatures $T$ the latter favors a ground state where translational symmetry is spontaneously broken. This manifests itself in a spatially inhomogeneous scalar condensate, or equivalently a coordinate dependent fermion mass. As exact solutions are available for both the condensate shape and the full Dirac spectrum \cite{Schnetz:2004,Thies:2003br}, one of those can be used as a paradigmatic example for mapping the Dirac equation in an inhomogeneous condensate onto photonic waveguide arrays.

The Gross-Neveu (GN) model was originally introduced in 1974 \cite{Gross:1974} as a toy model for quantum chromodynamics, that is, the theory of the strong interaction. It is a fermionic relativistic quantum field theory in 1+1 space-time dimensions, and describes $N$ species of massless Dirac fermions $\psi^{(n)}=\psi^{(n)}(x)$, with $n\in\{1,\ldots,N\}$, interacting with each other via a four-fermion interaction; $x^\mu=(t,{\rm x})$. The (massless) GN model is defined by the Lagrangian
\begin{equation}
 {\cal L}=\sum_{n=1}^N\bar\psi^{(n)} {\rm i}\slashed{\partial}\psi^{(n)} + \frac{1}{2}g^2\Bigl(\sum_{n=1}^N\bar\psi^{(n)}\psi^{(n)}\Bigr)^2 , \label{eq:Lorig}
\end{equation}
with two-component spinor fields $\psi^{(n)}=(\psi_1^{(n)},\psi_2^{(n)})^T$ and $\bar\psi^{(n)}=(\psi^{(n)})^\dag\gamma^0$. Here, we use natural units $c=\hbar=1$ and employ the shorthand notation $\slashed\partial=\gamma^\mu\partial_\mu=\gamma^0\partial_t+\gamma^1\partial_{\rm x}$, with $\gamma^0=\beta$ and $\gamma^1=\beta\alpha$ denoting the Dirac matrices in 1+1 dimensions, where $\mu$ runs from zero to one.
The coupling $g$ is dimensionless in 1+1 dimensions. 

The equation of motion of the fermion field $\psi^{(i)}$ follows from the Euler-Lagrange equation, yielding
\begin{equation}\label{eq:EoM}
  \Bigl({\rm i}\slashed{\partial} + g^2\sum_{n=1}^N\bar\psi^{(n)}\psi^{(n)}\Bigr)\psi^{(i)}=0\,.
\end{equation}
As the GN model is a relativistic QFT, and thus genuinely a multi-particle theory, the fermion fields are not single-particle wavefunctions but second quantized field operators featuring infinitely many positive and negative energy states. This implies that also the bilinear combination of the fermion fields inside the parentheses in Eq. \eqref{eq:EoM} is operator valued.

In the 't Hooft limit, defined by sending $N\to\infty$ while keeping $Ng^2={\rm const}.$, this bilinear combination can be replaced by its expectation value in the considered multi-fermion state \cite{Pausch:1991}. Corrections are parametrically suppressed by inverse powers of $N$, rendering this replacement exact. In this large-$N$ limit, the equation of motion~\eqref{eq:EoM} reduces to
\begin{equation}
  \bigl({\rm i}\slashed{\partial} - S(x)\bigr)\psi^{(i)}=0\,, \label{eq:EoMHF}
\end{equation}
with scalar potential $S(x)$ given by
\begin{equation}
  S(x)=-g^2\sum_{n=1}^N\big\langle\bar\psi^{(n)}\psi^{(n)}\big\rangle\,. \label{eq:SCC}
\end{equation}
Here, we focus on static configurations for which the expectation value $\big\langle\bar\psi^{(n)}\psi^{(n)}\big\rangle$ is time-independent. Equation \eqref{eq:EoMHF} corresponds to the Dirac equation describing fermions with a prescribed coordinate dependent mass $S(x)$. The difficulty in solving it arises from the self-consistency condition \eqref{eq:SCC}, that is, $S(x)$ itself is defined in terms of solutions of the Dirac equation \eqref{eq:EoMHF}. Aiming at the ground state of the system for given fermion density and temperature, one has to solve Eqs. \eqref{eq:EoMHF} and \eqref{eq:SCC} with infinitely many single particle states occupied: apart from all negative energy states, additional energy levels are to be populated until the prescribed value for the fermion density is reached. The infinite sum over the negative energy states is divergent and requires regularization. However, it turns out that all dependences on the regularization, as well as the bare coupling constant $g$ can be traded for the physical fermion mass $m$. The latter is the only parameter of the translationally invariant vacuum characterized by a filled Dirac sea and all positive energy levels empty \cite{Pausch:1991}. The physical fermion density in the system is also measured relative to the vacuum state. Multi-fermion bound states of $N_f\leq N$ fermions in excess to the vacuum are often referred to as \textit{baryons} \cite{Witten:1979kh} due to their analogy to baryons in hadron physics. Baryon number one is assigned to a state with $N_f=N$, and the fermion density in the system is conventionally parameterized by the baryon density $\rho$. The similarity of the model to the strong interactions has triggered a substantial amount of analytical and numerical studies in recent years \cite{deForcrand:2006zz,Karbstein:2006er,Dunne:2009zz,Basar:2010mu,Braun:2014fga,Pannullo:2019bfn,Narayanan:2020uqt,Lenz:2020bxk,Lenz:2020cuv,Stoll:2021ori}. 

\begin{figure}[b]
\includegraphics[width=0.9\columnwidth]{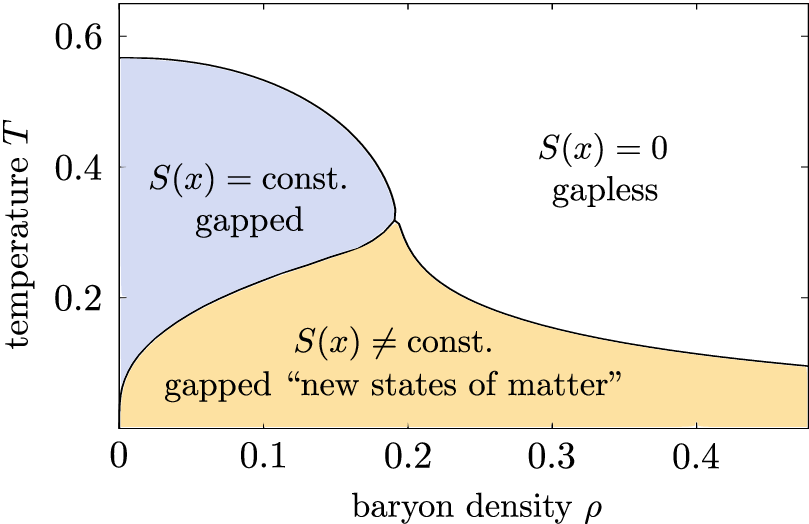}
\caption{\label{fig:fig1} Phase diagram of the theory. New states of matter characterized by $S(x)\neq{\rm const}.$ arise for small temperatures \cite{Thies:2003kk}.}
\end{figure}

In Fig. \ref{fig:fig1}, the phase diagram of the GN model in the $(\rho,T)$ plane is shown \cite{Thies:2003kk}. In the regions where $S(x)={\rm const}.$ and $S(x)=0$ the ground state of the system is translationally invariant. In the gapped phase, an effective mass for the fermions is spontaneously generated by chiral symmetry breaking, while they remain massless in the gapless phase. Finally, in the phase where the scalar condensate $S(x)$ exhibits an explicit dependence on $x$ the ground state breaks translational symmetry and is characterized by a coordinate-dependent fermion mass. This is the regime of new states of matter \cite{Thies:2003kk,Thies:2003br,Schnetz:2004}; cf. also \cite{Mertsching:1981,Machida:1984zz} for applications in condensed matter physics.

\begin{figure}[b]
\includegraphics[width=0.85\columnwidth]{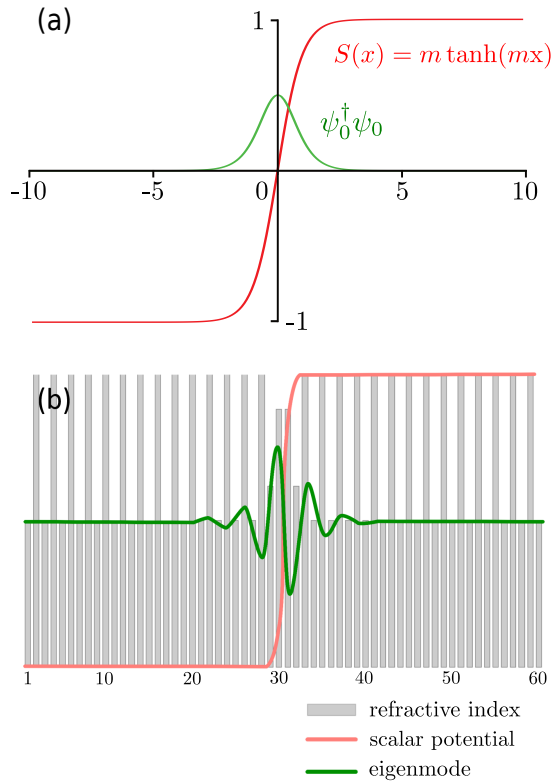}
\caption{\label{fig:fig2} (a) Scalar potential (red) and associated valence number density (green) in units where $m=1$. (b) Photonic emulation in an array of waveguides with alternating high and low refractive indices. The emulated scalar potential (red) gives rise to a localized mode in the center of the structure.}
\end{figure}

It can be shown that the single-{\it kink} scalar potential
\begin{equation}\label{eq:kink}
 S({\rm x})=m\tanh(m{\rm x})
\end{equation}
corresponds to one of the cases for which Eqs. \eqref{eq:EoMHF} and \eqref{eq:SCC} can be solved analytically \cite{Dashen:1975}. Equation~\eqref{eq:kink} is fully determined by the physical fermion mass $m$. It is characterized by a filled negative energy continuum and a valence level populated with $n\leq N$ fermions lying right in the middle of the energy gap separating the negative and positive energy continua; see Fig.~\ref{fig:fig2}(a) for an illustration. The total fermion number associated with this object is $N_f=n-N/2$ \cite{Karbstein:2007bg}. As it interpolates between asymptotic states with positive and negative physical fermion mass $m$ for ${\rm x}\to\pm\infty$, respectively, it amounts to a manifestly relativistic object which has no non-relativistic limit \cite{Pausch:1991}.

Now, the idea of our work is to emulate the Dirac equation by a waveguide array implementing a spatially inhomogeneous potential $S(x)$, which is a self-consistent solution of Eqs. \eqref{eq:EoMHF} and \eqref{eq:SCC}. In this manner, multi-fermion bound-state formation, which is closely related to the phenomenon of translational symmetry breaking \cite{Thies:2006ti}, can be directly probed. Such an experiment illustrates that photonic platforms have the potential to serve as a viable laboratory for the physics of relativistic self-interacting fermion systems; in this way,  a new and flexible tool to search for new states of matter in photonic experiments becomes available.

Equation \eqref{eq:EoMHF} can be rewritten as
\begin{equation}
  {\rm i}\partial_t\psi^{(i)}=-{\rm i}\alpha\partial_{\rm x}\psi^{(i)} + \beta S({\rm x})\psi^{(i)}\,. \label{eq:EoMstart}
\end{equation}
We choose $\alpha=\sigma_1$ and $\beta=\sigma_3$, with Pauli matrices $\sigma_i$, and discretize the spatial coordinate $\rm x$ on a lattice. Obviously, the off-diagonal matrix $\alpha$ couples the equations for the upper and lower spinor components. Following \cite{Longi:2010}, we decompose a one-dimensional lattice with lattice spacing $d$ into two independent sublattices of lattice spacing $2d$ to accommodate the two independent components of the spinor field: the upper component resides on the even sites, the lower component on the odd sites. In turn, the pair of lattice sites $2n$ and $2n-1$, with $n\in\mathbb{Z}$, is associated with the same spatial coordinate ${\rm x}$. When expressing the components of the spinor field as $\psi_1^{(i)}({\rm x},t)\to(-1)^n a_{2n}^{(i)}(t)$ and $\psi_2^{(i)}(x)\to{\rm i}(-1)^n a_{2n-1}^{(i)}(t)$ in terms of complex amplitudes $a_{\cal N}^{(i)}(t)$, one can discretize the scalar potential $S({\rm x})\to S({\cal N}d)$ with ${\cal N}=2n$ for the upper and ${\cal N}=2n-1$ for the lower spinor component, respectively. This results in the equation \cite{Longi:2010,Cannata:1990}
\begin{equation}
 {\rm i}\frac{\rm d}{{\rm dz}} a^{(i)}_{\cal N}=-\bigl(a^{(i)}_{{\cal N}+1}+a^{(i)}_{{\cal N}-1}\bigr)+S_{\cal N}(-1)^{\cal N}a^{(i)}_{\cal N}\,,
 \label{eq:discretizedEoM}
\end{equation}
where we introduced the dimensionless time ${\rm z}=t/(2d)$ and the dimensionless potential $S_{\cal N}=2d\,S({\cal N}d)$. The evolution of the amplitude $a^{(i)}_{\cal N}$ with $\rm z$ is coupled to the neighboring amplitudes  $a^{(i)}_{{\cal N}\pm1}$. Equation~\eqref{eq:discretizedEoM} is amenable to a waveguide implementation \cite{Dreisow:2010}. As each of the distinct fermion species $i\in\{1,\ldots,N\}$ fulfills the same equation, it suffices to simulate it for a single fermion species. The coordinate $\mathrm{z}$ can be mapped onto the longitudinal coordinate of the waveguide array, and $a^{(i)}_{\cal N}$ to the amplitude of the light coupled into the $\cal N$th waveguide. Here, it experiences the waveguide-specific index of refraction ${\cal S}_{\cal N}$, as shown in Fig.~\ref{fig:fig2}(b). Using our conventions for the Dirac matrices, the analytical solution for the valence spinor associated with the single kink potential~\eqref{eq:kink} is given by \cite{Dashen:1975,Pausch:1991}
\begin{equation}
\psi_0({\rm x})=\frac{\sqrt{m}\ {\rm e}^{-{\rm i}\varphi}}{2\cosh(m{\rm x})}\left(\begin{array}{r} 1 \\ {\rm i} \end{array}\right) \,,
\label{eq:psh_DHN}
\end{equation}
where $\varphi$ is an arbitrary global phase, which drops out in observables such as $\psi_0^\dag\psi_0$.

\begin{figure}[t]
\includegraphics[width=\columnwidth]{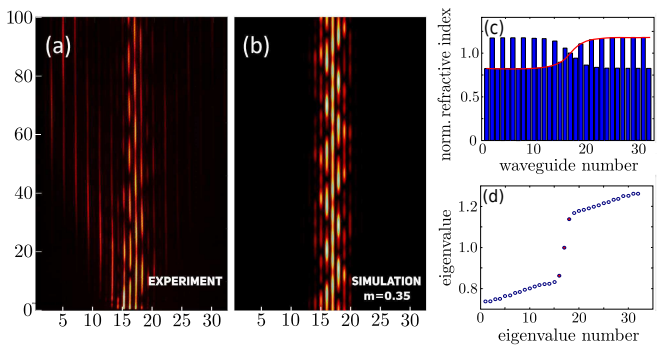}
\caption{\label{fig:fig3} (a) Experimental light dynamics in the photonic structure. Clearly, the beam is localized, which emulates the valence state in the GN model. (b) Numerical confirmation of the experimental results, obtained by integrating Eq.~\eqref{eq:discretizedEoM}. (c) Normalized refractive index of the individual waveguides (blue) and corresponding scalar potential (red). (d) Eigenvalue diagram of the implemented photonic structure. Clearly, the GN valence state at the kink is visible in the middle of the gap (in red). Besides, the photonic structure supports two additional localized states at the edge of the band gap (also in red).}
\end{figure}

For our experiments, we fabricate various waveguide lattices using the direct-laser writing technology \cite{Szameit:2010}. The length of the waveguides is $l=100\,{\rm mm}$, the lattice constant is $d=16\,\upmu{\rm m}$, and the average refractive index of each full lattice is $\delta n = 6\times 10^{-4}$. In order to implement the last term in Eq. \eqref{eq:discretizedEoM}, the two sublattices are realized by fabricating an alternating sequence of waveguides with high and low refractive index change \cite{Dreisow:2010}. This tuning is accomplished by varying the ratio between the writing velocities of adjacent waveguides in the sublattices; the magnitude of the writing velocity difference is proportional to the scalar potential $S_{\cal N}$. Importantly, changing the writing speed leaves the intersite hopping of $\kappa = 0.14\,{\rm mm}^{-1}$ essentially unchanged \cite{Heinrich:2014}. A fluorescence microscopy technique \cite{Szameit2007} enables us to map the flow of light from the top of the sample and, thus, to visualize the spinor wave packet evolution. The array is excited by a broad Gaussian beam at a wavelength of $\lambda = 633\,{\rm nm}$ with a spot size of $\sim 80\,\upmu{\rm m}$ in the transverse direction, covering approximately 5 waveguides. A representative example of dynamics in such a fabricated waveguide structure is shown in Fig.~\ref{fig:fig3}. The refractive index detuning between the sublattices far away from the kink at the center of the structure -- that is, at $S_{\cal N}(\pm\infty)$ -- is $\Delta n \approx 2\times 10^{-4}$. Upon excitation in the center of the array, i.e., directly at the kink, clearly a bound dynamics is observed in the form of an oscillatory motion of the light beam (see Fig.~\ref{fig:fig3}(a)) matching the analytical predictions for the valence level of the kink potential~\eqref{eq:kink}. Our observation is supported by a numerical integration of \Eqref{eq:discretizedEoM}, that yields dynamics that resembles our experimental data (see Fig.~\ref{fig:fig3}(b)). The implemented (normalized) refractive index distribution of the waveguide lattice is shown in Fig.~\ref{fig:fig3}(c), emulating the scalar potential~\eqref{eq:kink} in the center of the structure. A plot of the respective eigenvalues is shown in Fig.~\ref{fig:fig3}(d), showing the mid-gap state~\eqref{eq:psh_DHN} that is localized at the kink.

\begin{figure}[b]
\includegraphics[width=\columnwidth]{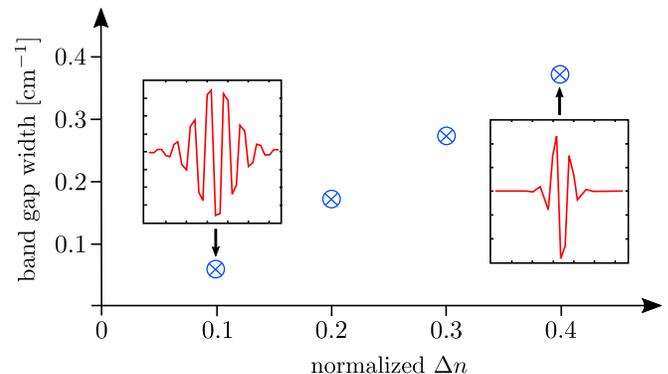}
\caption{\label{fig:fig4} Experimentally obtained approximate width of the band gap as a function of the refractive index detuning. The insets show examples of the associated eigenmodes characterizing the valence spinor~\eqref{eq:psh_DHN} discretized in the waveguide array as a function of the waveguide number; cf. also Fig.~\ref{fig:fig2} (b).}
\end{figure}

Moreover, the eigenvalue diagram~\ref{fig:fig3}(d) shows that not only the mid-gap bound state as predicted for the GN valence state is emerging, but that also the two states at the inner edges of the bands slightly penetrate into gap and, hence, become localized bound states as well. This behavior is a result of the intrinsic lattice discretization and, hence, a characteristic feature of our photonic emulator. When launching light into the center of the lattice, all three states are excited; as a result, one observes a beating between the states. However, this beating is spatially confined in the very vicinity of the kink. Interestingly, the beating length $l_\mathrm{B}$ can be used to estimate the width of the band gap in the experimental system, as it is connected to the energy difference $\Delta E$ of the states by $l_\mathrm{B} = 2\pi/\Delta E$ \cite{Yariv:1989}. As the other localized states reside close to the edge of the band gap, $\Delta E$ is a measure for the band gap width. To explore this feature, we conducted several experiments similar to that in Fig.~\ref{fig:fig3} with different normalized refractive index detuning $\Delta n$ at the edges of the waveguide lattices (i.e., at $S_{\cal N}(\pm\infty)$). In Fig.~\ref{fig:fig4}, we plot the experimentally determined gap widths as a function of $\Delta n$. Evidently, the band gap width increases almost linearly for increasing detuning. Equation~\eqref{eq:kink} implies that the width of the band gap is directly related to the fermion mass $m$, which also determines the spatial localization of the valence spinor~\eqref{eq:psh_DHN}. Hence, our emulator allows for an indirect determination of both the fermion mass and the spatial localization of the valence spinor (see the insets in Fig.~\ref{fig:fig4}), via observing the oscillation period of the light wave.

In conclusion, we reported an implementation of a photonic emulator for phenomena related to bound-state formation and translational symmetry breaking in the ground state of the large-$N$ Gross-Neveu model. Our results suggest that waveguide optics could provide an experimentally accessible classical emulator to test complex predictions concerning the spontaneous breaking of translational invariance and the formation of spatially inhomogeneous phases in the phase diagram of the theory.
A particularly interesting future application of our photonic emulator will be the study of multi-fermion bound state formation in quantum field theories in the non-relativistic limit. Here, the Dirac sea involving infinitely many filled negative-energy states can be integrated out. This results in a {\it no-sea} effective field theory featuring fermion fields of manifestly positive energy only \cite{Karbstein:2007be}, for the study of which our photonic emulator should be ideally suited. The no-sea analogue of the Gross-Neveu model as well as many other 1+1 dimensional field-theories with four-fermion interactions can even be solved analytically which provides various benchmark solutions for such studies \cite{Lee:1975tx}.
In particular for phenomena in 2D and 3D settings, where the search for new states of matter is extremely challenging, photonics may provide a new promising route for exploring these phenomena experimentally.

The authors would like to thank C.~Otto for preparing the high-quality fused silica samples employed in this work. This work has been carried out within the framework of the ACP Explore project ``Enlightning New States of Matter'' of the Abbe Center of Photonics (ACP). AS acknowledges financial support from the European Research Council (grant EPIQUS), the Alfried Krupp von Bohlen and Halbach foundation, and the Deutsche Forschungsgemeinschaft (grants SCHE 612/6-1, SZ 276/12-1, BL 574/13-1, SZ 276/15-1, SZ 276/20-1, and SFB 1477 "Light-Matter Interactions at Interfaces", project number 441234705). HG acknowledges funding by the Deutsche Forschungsgemeinschaft und grant No. 398579334 (Gi 328/9-1).

\end{document}